\documentclass[preprint,superscriptaddress]{revtex4-1}

\usepackage{graphicx}
\usepackage{epstopdf}
\usepackage[ansinew]{inputenc}
\usepackage{array}
\usepackage{color}
\usepackage{amsmath}
\usepackage{amsxtra}
\usepackage{amstext}
\usepackage{amssymb}
\usepackage{latexsym}
\usepackage{dsfont}
\usepackage{soul}

\begin{document}
\title{Intrinsic disorder in graphene on transition metal dichalcogenide heterostructures}
\author{Matthew Yankowitz}
\affiliation{Physics Department, University of Arizona, Tucson, AZ 85721, USA}
\author{Stefano Larentis}
\affiliation{Microelectronics Research Center, The University of Texas at Austin, Austin, TX 
78758, USA}
\author{Kyounghwan Kim}
\affiliation{Microelectronics Research Center, The University of Texas at Austin, Austin, TX 
78758, USA}
\author{Jiamin Xue}
\affiliation{Microelectronics Research Center, The University of Texas at Austin, Austin, TX 
78758, USA}
\author{Devin McKenzie}
\affiliation{Physics Department, University of Arizona, Tucson, AZ 85721, USA}
\author{Shengqiang Huang}
\affiliation{Physics Department, University of Arizona, Tucson, AZ 85721, USA}
\author{Marina Paggen}
\affiliation{Physics Department, University of Arizona, Tucson, AZ 85721, USA}
\affiliation{Physics Department, University of Texas at El Paso, El Paso, TX 79968, USA}
\author{Mazhar N. Ali}
\affiliation{Department of Chemistry, Princeton University, Princeton, New Jersey 08544, USA}
\author{Robert J. Cava}
\affiliation{Department of Chemistry, Princeton University, Princeton, New Jersey 08544, USA}
\author{Emanuel Tutuc}
\affiliation{Microelectronics Research Center, The University of Texas at Austin, Austin, TX 
78758, USA}
\author{Brian J. LeRoy}
\email{leroy@physics.arizona.edu}
\affiliation{Physics Department, University of Arizona, Tucson, AZ 85721, USA}
\date{\today}

\begin{abstract}
The electronic properties of two-dimensional materials such as graphene are extremely sensitive to their environment, especially the underlying substrate. Planar van der Waals bonded substrates such as hexagonal boron nitride (hBN) have been shown to greatly improve the electrical performance of graphene devices by reducing topographic variations and charge fluctuations compared to amorphous insulating substrates~\cite{Yankowitz2014,Dean2010,Xue2011,Decker2011}. Semiconducting transition metal dichalchogenides (TMDs) are another family of van der Waals bonded materials that have recently received interest as alternative substrates to hBN for graphene~\cite{Kretinin2014,Larentis2014,Tan2014} as well as for components in novel graphene-based device heterostructures~\cite{Britnell2012,Georgiou2013,Yu2013a,Moriya2014,Britnell2013,Yu2013b,Bertolazzi2013,Roy2013}. Additionally, their semiconducting nature permits dynamic gate voltage control over the interaction strength with graphene~\cite{Lu2014b}. Through local scanning probe measurements we find that crystalline defects intrinsic to TMDs induce scattering in graphene which results in significant degradation of the heterostructure quality, particularly compared to similar graphene on hBN devices.
\end{abstract}

\maketitle
Since the isolation of graphene in 2004, considerable effort has been put into finding the best substrates, both from a device standpoint and for inducing novel physical phenomena. Perhaps the most common substrate, SiO$_2$, causes out-of-plane ripples and locally dopes the graphene due to trapped charged impurities~\cite{DasSarma2011}. Suspended graphene devices, fabricated by etching away the SiO$_2$ layer, offer the best intrinsic graphene quality, although their fabrication is far too challenging to scale to industrial levels. Hexagonal boron nitride has emerged as a very promising substrate; as an insulating crystal it not only flattens the graphene but screens underlying charge impurities from the base substrate~\cite{Yankowitz2014,Dean2010,Xue2011,Decker2011}. Careful device fabrication techniques can yield devices of graphene quality nearing that of suspended graphene, and these heterostructures are more friendly for industrial scaling. Since hBN has a similar lattice constant to graphene, when the two lattices are in near perfect alignment interactions between the crystals strongly renormalize the graphene band structure~\cite{Yankowitz2012,Yankowitz2014}. This opens an avenue for the study of new phenomena, such as the Hofstadter quantization, and also provides a route to make graphene insulating~\cite{Yankowitz2014}. However, it may not be ideal for large scale graphene device applications where fabrication leads to a random alignment between the crystals, and the intrinsic graphene band structure needs to be preserved.

Recently, transition metal dichalcogenides have made a strong resurgence in materials research, as these crystals can be exfoliated to atomic scale thicknesses and stacked via van der Waals interactions similarly to graphene and hBN~\cite{Wang2012,Geim2013}. A subset of the TMDs exhibit similar semiconducting behavior, with indirect band gaps in bulk ranging from $\sim$ 1 - 1.4 eV~\cite{Wang2012}. Na\"{\i}vely, these materials, when insulating, should offer comparable quality to hBN as substrates for graphene, but without the possibility of band structure modification due to their considerably different lattice constants~~\cite{Wilson1969,Lu2014b}. Additionally, they offer the potential for the study of new physical phenomena (for example, potential spin-orbit coupling induced in the graphene layer due to the heavy metal atoms of the TMD~\cite{Kretinin2014}). From a device standpoint, there are numerous potential applications involving heterostructures between graphene and TMDs; for example, as tunneling transistors~\cite{Britnell2012,Georgiou2013,Yu2013a,Moriya2014}, highly efficient flexible photovoltaic devices~\cite{Britnell2013,Yu2013b}, or nonvolatile memory cells~\cite{Bertolazzi2013,Roy2013}. Unfortunately, graphene on TMD devices have thus far been of significantly lower mobility than comparable hBN devices~\cite{Kretinin2014,Larentis2014,Tan2014}, and a local understanding of this behavior is lacking. In this Letter, we show via local scanning probe measurements that graphene on TMD devices suffer an unavoidable degradation in electronic quality due to intrinsic defects in the TMD crystals. 

\begin{figure}[t]
\includegraphics[width=8.5cm]{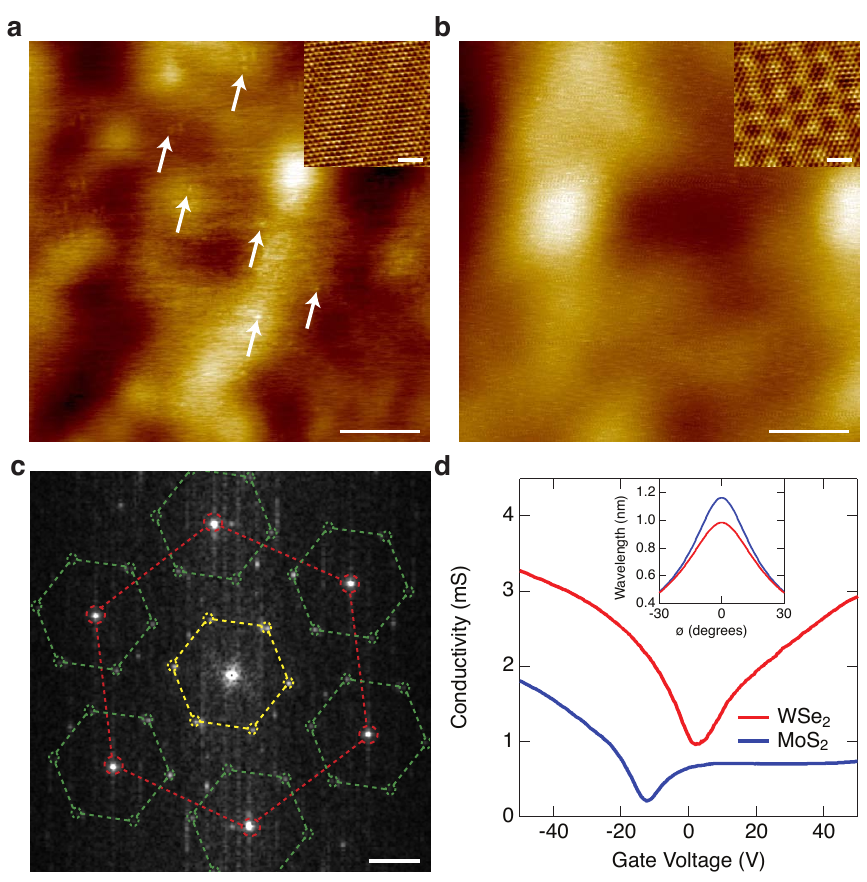} 
\caption{Topography and electrical transport. (a) Topography of graphene on MoS$_2$. Arrows mark some of the defects buried in the MoS$_2$. Inset: Atomic resolution with a moir\'e pattern of 0.65 nm. (b) Same as (a) for WSe$_2$. No defects are visible. Inset: Atomic resolution with a moir\'e pattern of 0.91 nm. For (a) and (b), the scale bar for the main figure is 10 nm and for the insets it is 1 nm. Typical imaging parameters are sample voltages between V$_s$ = 0.1 V and 0.3 V and tunnel currents between I$_t$ = 100 pA and 300 pA. (c) Fourier transform of (a). The atomic lattice is highlighted with a red hexagon, and the moir\'e with a yellow hexagon. The moir\'e points surrounding the atomic lattice are highlighted with green hexagons. The scale bar is 10 nm$^{-1}$. (d) Global conductance measurements of graphene on MoS$_2$ and WSe$_2$ devices. The MoS$_2$ becomes conducting at gate voltages just above 0 V, after which point the device exhibits negative compressibility. The WSe$_2$ crystal is always biased within the band gap. Inset: Dispersion of possible moir\'e wavelengths for graphene on MoS$_2$ and WSe$_2$ as a function of the relative rotation between the lattices. }
\newpage
\label{fig:topography}
\end{figure}

We study graphene on substrates belonging to the MX$_2$ family, where M is a transition metal (Mo, W) and X is a chalcogen atom (S, Se, Te). The specific TMDs examined here are MoS$_2$, WS$_2$, WSe$_2$, and MoTe$_2$. We have also examined graphene on SnS$_2$, which is not technically a TMD but shares the same crystal structure and is also a semiconductor~\cite{Greenaway1965,Sharp2006}. Figs. ~\ref{fig:topography}(a) and (b) show topography images of graphene on MoS$_2$ and WSe$_2$ obtained via scanning tunneling microscopy (STM). Numerous defects can be observed in the MoS$_2$ sample. The graphene lattice is continuous over these defects, indicating the defects reside in the MoS$_2$. The strength and appearance of these defects can be tuned by the sample and gate voltages, and their appearance in other TMDs is found to depend on those factors as well as crystal thickness. The graphene on MoTe$_2$ and SnS$_2$ samples exhibit significantly worse topography, marked by islands of well-adhered regions surrounded by significantly rougher regions where the graphene did not appear to be adhered to the TMD, suggesting that these crystals are not stable enough in air to allow consistently good adhesion with the graphene~\cite{Geim2013} (see Supplementary Information for further discussion of topography observations).

The insets of Figs. ~\ref{fig:topography}(a) and (b) show atomic resolution images of both samples, and additionally display hexagonal superlattices due to the interference pattern formed by the graphene and TMD lattices. As is the case for graphene on hBN, a moir\'e pattern is expected to form between the graphene and TMD lattices due to their relative rotation $\phi$ and lattice mismatch $\delta$.  The moir\'e wavelength is
$$\lambda = \frac{(1+\delta)a}{\sqrt{2(1+\delta)(1-\cos\phi)+\delta^2}}$$
where $a$ is the graphene lattice constant~\cite{Yankowitz2012}.  Because the lattice mismatch is much larger between graphene and TMDs than graphene and hBN, the range of possible moir\'e wavelengths is much smaller. The inset of Fig. ~\ref{fig:topography}(d) shows the dispersion of moir\'e wavelengths as a function of angle $\phi$ for graphene on MoS$_2$ and WSe$_2$ (the dispersions for the other TMDs studied here are very similar). The lattice constants of the TMDs in this study range from 3.15 to 3.64 \AA~\cite{Wilson1969,Greenaway1965}, and thus the possible moir\'e wavelengths are on the order of 0.5 nm to just over 1 nm. As a result of these short moir\'e wavelengths, no modification of the graphene band structure is expected at low energies~\cite{Yankowitz2012}. Figure ~\ref{fig:topography}(c) shows a Fourier transform of the inset of Fig. ~\ref{fig:topography}(a), with the superlattice (yellow/green) and graphene lattice (red) hexagons highlighted.

\begin{figure}[t]
\includegraphics[width=8.5cm]{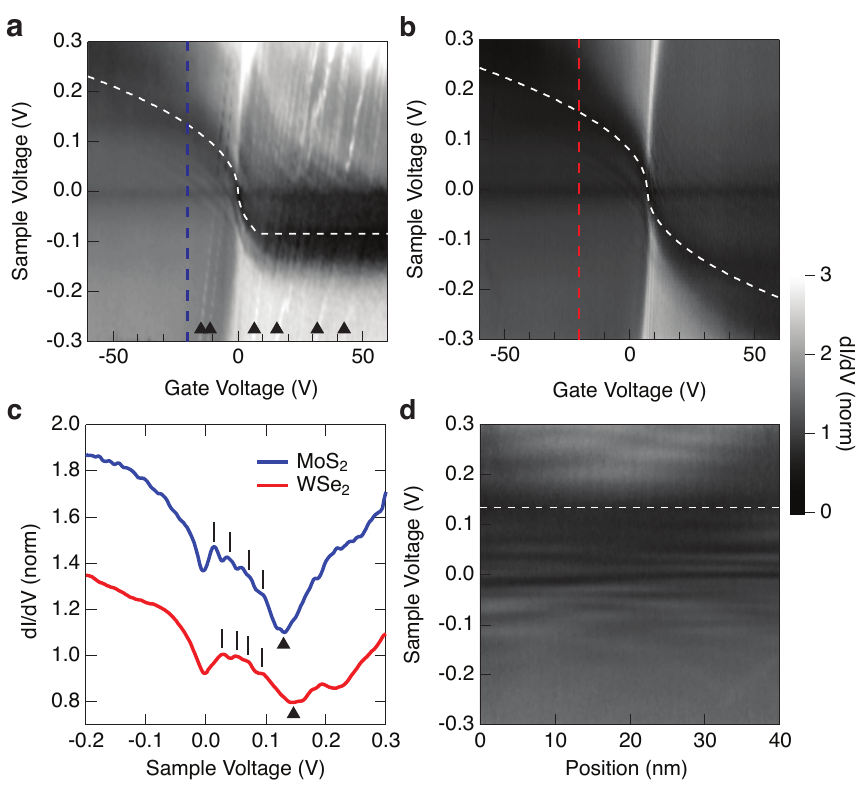} 
\caption{Local spectroscopy. (a) Normalized (dI/dV)/(I/V) spectroscopy as a function of gate voltage for graphene on MoS$_2$. The Dirac point is highlighted with a dotted white line. The Dirac point remains nearly stationary with gate voltage above $\sim$+10 V as the MoS$_2$ becomes conducting and the charge density in the graphene is constant. States due to charging of defects in MoS$_2$ are marked with black arrows. (b) Same as (a) for graphene on WSe$_2$. In both (a) and (b), extra peaks of varying strength due to scattering surround the Dirac point at both positive and negative energies. (c) Cuts of (a) and (b) (marked by the dashed vertical lines) at V$_g$ = -20 V. For both, the Dirac point is around V$_s$ = +0.15V (marked with black arrows). Extra states due to scattering are marked with tick marks. (d) Normalized dI/dV spectroscopy as a function of position in graphene on MoTe$_2$ at V$_g$ = -9 V. The approximate position of the Dirac point is marked with a dotted white line. All other resonances are attributed to intravalley scattering, and their energy spacing varies randomly with position.}
\newpage
\label{fig:spectroscopy}
\end{figure}

In addition to their similar crystal structure, these TMDs also share similar electronic properties. They are indirect gap semiconductors with band gaps in bulk ranging from about 1.0 eV to 1.4 eV ~\cite{Wang2012} (and 2.2 eV for SnS$_2$~\cite{Sharp2006}). In proximity to graphene, the difference in energy between the graphene work function and the electron affinity of the TMD determines the relative band alignment. The threshold gate voltage at which the TMD begins to conduct depends on this band alignment as well as the band bending related to the TMD crystal thickness~\cite{Larentis2014}. For comparably thin crystals ($\sim$5 - 15 nm), we observe that the Dirac point is aligned very near the conduction bands of MoS$_2$, MoTe$_2$ and SnS$_2$, and closer to the middle of the gap for WS$_2$ and WSe$_2$.  Figure ~\ref{fig:topography}(d) shows global conductance measurements at 4.2 K for graphene on MoS$_2$ and WSe$_2$. The graphene on WSe$_2$ device is nearly symmetric around charge neutrality (indicating the WSe$_2$ always remains insulating), whereas the MoS$_2$ device has a saturating conductance at positive gate voltages as the MoS$_2$ becomes conducting. We find field-effect mobilities of 5,000 - 10,000 cm$^{2}$/Vs in these devices, which is over an order of magnitude lower than of comparable graphene on hBN devices. We restrict our study to thin TMD substrates as devices with thicker crystals become strongly hysteretic with gate voltage.

Figures ~\ref{fig:spectroscopy}(a) and (b) show normalized (dI/dV)/(I/V) spectroscopy measurements as a function of gate voltage for graphene on MoS$_2$ and WSe$_2$. For MoS$_2$, the Dirac point (marked with a white dotted line) essentially stops moving in energy once the gate voltage is large enough to induce charge carriers in the TMD. In some samples we have additionally observed the Dirac point move with positive dV$_s$/dV$_g$ when the MoS$_2$ is conducting, indicative of the negative compressibility of the system~\cite{Larentis2014} (see Supplementary Information). For graphene on WSe$_2$, the Dirac point moves with the standard square root of gate voltage dispersion expected for graphene on an insulating substrate~\cite{Zhang2008}, indicating the WSe$_2$ substrate is never biased to the valence or conduction bands. In this case, the WSe$_2$ should behave very similarly to an hBN substrate, flattening the graphene and screening charged impurities in the underlying SiO$_2$. For all the graphene on TMD devices, the movement of the Dirac point is well fit by a Fermi velocity of 0.95 $\pm$ 0.05 x 10$^{6}$ m/s.

\begin{figure}[t]
\includegraphics[width=8.5cm]{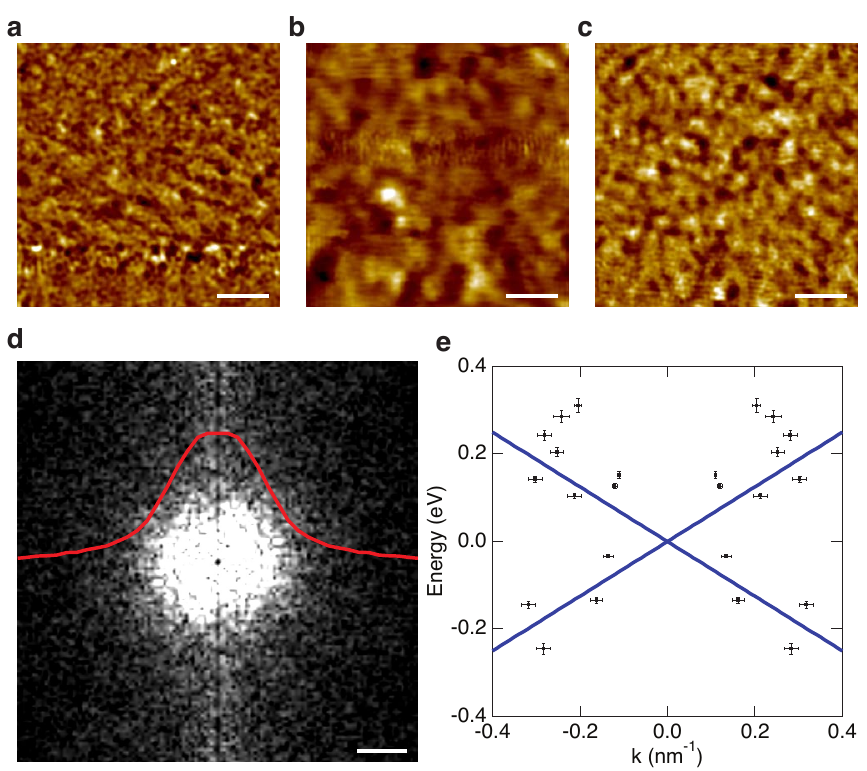} 
\caption{Intravalley scattering and extracted graphene dispersion. (a) - (c) Maps of dI/dV spectroscopy for graphene on WSe$_2$ at V$_g$ = -60 V, -5 V, and +60 V, respectively. The local charge neutrality point is around V$_g$ = -15 V. Each map is taken with V$_s$ = 50 mV and I$_t$ = 150 pA. The coherent structure is due to intravalley scattering, and becomes shorter wavelength at gate voltages further from the Dirac point. The scale bars in (a)-(c) are 50 nm.  (d) Fourier transform of (c), exhibiting a disk-like feature at the center corresponding to the scattering wave vector. The circular average is plotted in red. The scale bar is 0.2 nm$^{-1}$. (e) Graphene energy versus momentum dispersion extracted from maps similar to those shown in (a) - (c). The wave vector $k$ is extracted from the half-width at half-maximum of the Lorentzian fit of the Fourier transforms of each map. The energy is determined by converting the sample voltage of each measurement to energy from the Dirac point using $v_F$ = 0.95 x 10$^{6}$ m/s. The solid blue lines plot the dispersion $E = \hbar v_F k$. The error in $k$ represents the uncertainty of the Lorentzian fit, and the error in $E$ represents the uncertainty in identifying the Fermi velocity.}
\newpage
\label{fig:dispersion}
\end{figure}

In addition to the usual spectroscopic features of graphene on an insulating substrate, we find there are extra, unexpected resonances in the dI/dV spectroscopy surrounding and moving roughly in parallel with the Dirac point. These features are ubiquitous amongst all MX$_2$ substrates examined in this study. Fig. ~\ref{fig:spectroscopy}(c) shows a line cuts of Figs. ~\ref{fig:spectroscopy}(a) and (b) at V$_g$ = -20 V, indicating clearly the presence of these extra states (marked with black ticks). We attribute these features to electronic scattering in graphene from point and line defects in the MX$_2$ substrates. We have considered other origins for the extra states we observe, such as new features of the band structure due to interactions with the  substrate, or the excitation of phonons in the substrate by electrons tunneling into the graphene. Such features should have well-defined energies which do not vary spatially or between different samples, depending on the exact nature of their origin. To test this hypothesis we take line maps of normalized dI/dV spectroscopy, as shown in Fig. ~\ref{fig:spectroscopy}(d) for graphene on MoTe$_2$. We see clearly that the energy spacing of these states varies spatially, and in some cases the states split or merge. While this behavior is inconsistent with the alternative explanations considered above, it is consistent with extra resonances resulting from intratravalley scattering in graphene from multiple point sources.

To further investigate the intravalley scattering in graphene on TMD heterostructures, we take large area dI/dV maps of graphene on WSe$_2$ at various gate voltages (see Figs. ~\ref{fig:dispersion}(a) - (c) for three examples). The coherent features in the maps change size with gate voltage, characteristic of intravalley scattering from the Coulomb potentials of buried defects. Fig. ~\ref{fig:dispersion}(d) shows the Fourier transform of Fig. ~\ref{fig:dispersion}(c). The disk at the center of the image arises from $2k$ scattering across a single graphene Dirac cone.  For each map, we take a circular average of the Fourier transform and extract the corresponding wave vector $k$ as the half-width at half-maximum of the best-fit Lorentzian. These wave vectors are plotted as a function of energy relative to the Dirac point in Fig. ~\ref{fig:dispersion}(e). The resulting dispersion is in good agreement with a Fermi velocity of $\sim$0.95 x 10$^{6}$ m/s.

\begin{figure}[t]
\includegraphics[width=8.5cm]{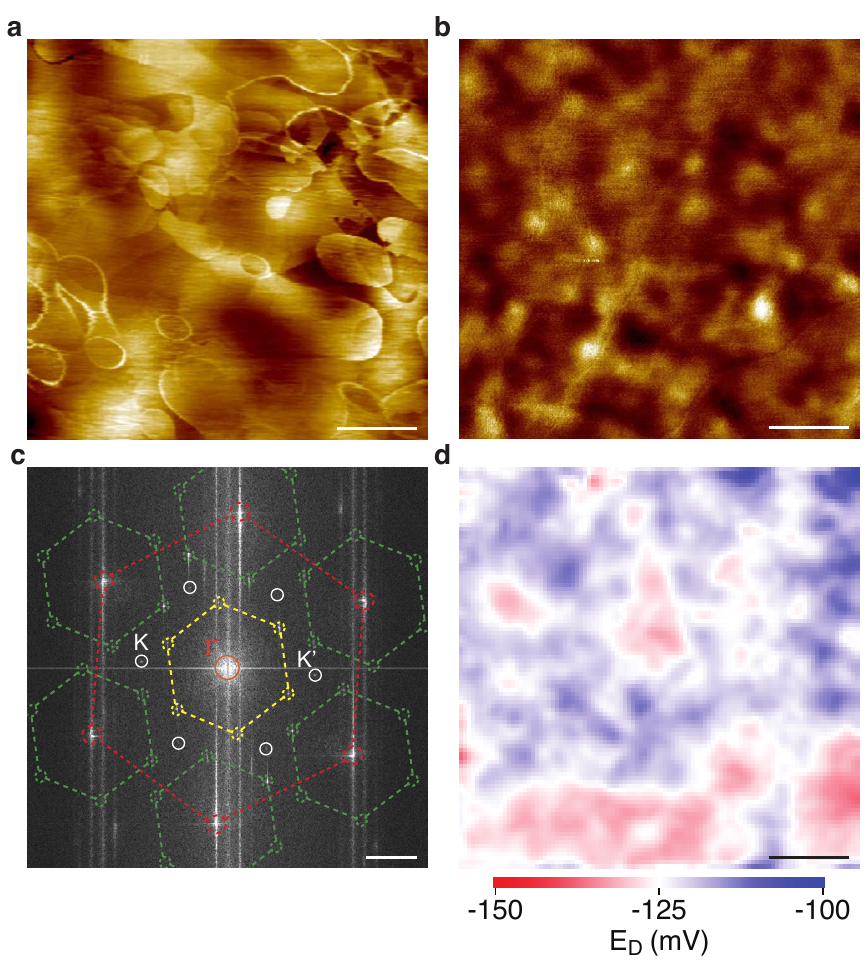} 
\caption{Defect scattering in graphene on TMDs. (a) dI/dV spectroscopy map of graphene on MoS$_2$. The numerous ring-like features are attributed to the charging of defect states by the STM tip. The map is taken with V$_s$ = 0.025 V, I$_t$ = 150 pA, and V$_g$ = 0 V. (b) Similar map of graphene on WSe$_2$. The faint wandering line features running through the map are attributed to line dislocations in the WSe$_2$ crystal. The map is taken with V$_s$ = 0.05 V, I$_t$ = 150 pA, and V$_g$ = +50 V. The scale bar is 10 nm in (a) and 20 nm in (b). (c) Fourier transform of an atomically resolved dI/dV map of graphene on MoS$_2$. In addition to the superlattice (yellow/green hexagons), atomic lattice (red hexagon), and long-wavelength intravalley scattering (red circle), there are also weak resonances due to intervalley scattering at the graphene K and K' points (white circles). The scale bar is 10 nm$^{-1}$. (d) Spatially resolved map of the Dirac point energy for graphene on WSe$_2$ at V$_g$ = 0 V. The scale bar is 50 nm.}
\newpage
\label{fig:defects}
\end{figure}

Fig. ~\ref{fig:defects}(a) shows a similar dI/dV map of graphene on MoS$_2$, which exhibits numerous ring features due to the charging or discharging of defect states resulting from the hybridization of MoS$_2$ defects with graphene (similar to those seen in artificial impurities on graphene~\cite{Brar2011}). The charging rings are the same features which run oppositely of the Dirac point in the gate map (features marked with black arrows in Fig. ~\ref{fig:spectroscopy}(a)). The size and nature of the rings can be tuned with back gate and sample voltage (similar ring structures have previously been observed in bare TMDs, but lacked this degree of tunability~\cite{Heckl1991,Magonov1994}). Of all the TMDs studied, these charging rings are unique to MoS$_2$ (both naturally occurring and synthetic). While no ring features are observed in graphene on WSe$_2$, there are numerous weak wandering line features which run through high resolution dI/dV maps (Fig. ~\ref{fig:defects}(b)). Since the graphene lattice is smooth over these line features and they are independent of sample and gate voltage, we attribute these to line dislocations intrinsic to the WSe$_2$ crystals. Similar features are observed in graphene on WS$_2$ as well. Fig. ~\ref{fig:defects}(c) shows the Fourier transform of an atomically resolved dI/dV map of graphene on MoS$_2$. The Fourier transform exhibits the usual resonances due to the superlattice (yellow/green hexagons) and atomic lattice (yellow hexagon), as well as the long-wavelength intravalley scattering (red circle). In addition, we observe weak resonances at the K and K' points of graphene (white circles). These resonances are indicative of intervalley scattering in graphene, which result from the presence of the atomic-scale point and line defects in the TMD substrates (for comparison, no such scattering is observed in clean graphene on hBN devices).

To address the possible influence of these defects on the charge environment in graphene, we take spatially resolved maps of the Dirac point energy (see Fig.~\ref{fig:defects}(d) for graphene on WSe$_2$). By converting the Dirac point energy $E_D$ to charge carrier density $n$ via $n = (1/\pi)(E_D/\hbar v_F)^{2}$, we find charge fluctuations of $\delta n \sim$ 1.4 $\pm$ 0.2 x 10$^{11}$ cm$^{-2}$ for graphene on MoS$_2$, WS$_2$, and WSe$_2$. While a few times better than the fluctuations observed in graphene on SiO$_2$~\cite{Zhang2009}, we find these to be around an order of magnitude larger than in comparable graphene on hBN devices~\cite{Xue2011,Decker2011}. These charge fluctuations are consistent with the defect density observed in bare MoS$_2$~\cite{Lu2014a}, suggesting that while the TMDs may be screening trapped charges in the SiO$_2$ substrate, their own intrinsic defects still create significant static charge disorder in the graphene. Even when the bottom few layers of the MoS$_2$ substrate are conducting and thus fully screening the SiO$_2$ interface (i.e. at large positive gate voltages), the observed charge fluctuations are only reduced by about a factor of two, which is still considerably larger than those observed with hBN. This further implies that the fluctuations are primarily due to TMD defects (the observed reduction is likely due to enhanced screening from the conducting MoS$_2$ layers). Finally, we do not observe any significant dependence on anneal temperature (between 150 $^{\circ}$C and 300 $^{\circ}$C), suggesting the defects in the TMD substrates are intrinsic to these crystals and are therefore unavoidable with current synthesis techniques.

Contrary to prior reports~\cite{Lu2014b}, we consistently observe a lower electronic quality of graphene on TMD devices than those using hBN. The dirtier local charge environment and prevalence of scattering is consistent with the lower mobility in our devices as well as those from prior reports~\cite{Kretinin2014,Larentis2014,Tan2014}. As heterostructures of graphene and TMDs grow quickly in popularity, it is critical to understand their intrinsic limitations. Unless new methods are developed for reducing defects in TMD crystals, the quality of these heterostructures will continue to be inferior to those of graphene on hBN.   

\section*{Methods}
Samples are fabricated by transferring either exfoliated or CVD-grown graphene flakes onto either naturally occurring (MoS$_2$) or synthetic (MoS$_2$, WS$_2$, WSe$_2$, MoTe$_2$, SnS$_2$) TMD flakes. TMD flakes are exfoliated directly onto a Si substrate capped with 285 nm of thermally grown SiO$_2$. Exfoliated graphene samples are transferred using the wet transfer technique utilized in Ref. ~\cite{Larentis2014}. We have not observed any difference in STM measurements between samples using exfoliated or CVD graphene, or between naturally occurring or synthetic MoS$_2$, so these distinctions are ignored. Samples are annealed in vacuum below $10^{-5}$ mbar at 300 $^{\circ}$C for MoS$_2$, WS$_2$, and WSe$_2$ (or 150 $^{\circ}$C where otherwise noted), and 250 $^{\circ}$C for MoTe$_2$ and SnS$_2$.

Naturally occurring bulk MoS$_2$ crystals were purchased from SPI. Synthetic MoS$_2$ crystals were purchased from 2D Semiconductors. WSe$_2$ crystals were purchased from Nanoscience Instruments. 2H-WS$_2$ crystals were grown by the direct vapor transport method of Ref.~\cite{Agarwal1979}. $\alpha$-MoTe$_2$ crystals were synthesized following the method of Ref.~\cite{AlHilli1972}. SnS$_2$ crystals were synthesized following the method of Ref.~\cite{Sharp2006}.

All the STM measurements were performed in ultrahigh vacuum at a temperature of 4.5 K.  dI/dV spectroscopy measurements were acquired by turning off the feedback circuit and adding a small (5-10 mV) a.c. voltage at 563 Hz to the sample voltage. The current was measured by lock-in detection. 

\section*{Acknowledgements}
The authors are thankful to Oliver Monti, David Racke, Laura Sharp, David Soltz, and Bruce Parkinson for supplying the SnS$_2$ crystals, as well as Allan H. MacDonald for valuable theoretical discussions.

M.Y., S.H. and B.J.L. were supported by NSF Career Award No. DMR-0953784.  S.L., K.K., J.X. and E.T. were supported by the Nanoelectronics Research Initiative and Intel.  D.M. was supported by the NSF REU program award No. PHY-1156753.

\section*{Author contributions}
M.Y. and B.J.L. performed the STM experiments. M.Y., S.L., K.K., J.X., and D.M. fabricated the devices. S.H. and B.J.L grew the CVD graphene. M.P. synthesized the bulk WS$_2$ crystal. M.N.A. and R.J.C. synthesized the bulk MoTe$_2$ crystal.  E.T. and B.J.L. conceived and provided advice on the experiments. M.Y. and B.J.L. wrote the manuscript with input from all authors.

\newpage

\section*{Supplementary Information}

\section{Topography of Graphene on TMDs}

In addition to graphene on MoS$_2$ and WSe$_2$, which are the focus of the main text, we have also examined graphene on WS$_2$, MoTe$_2$, and SnS$_2$. Figs.~\ref{fig:topography}(a), (c) and (d) show representative topography measurements of each of those samples. Graphene on WS$_2$ is topographically similar to MoS$_2$ and WSe$_2$. There are visible defects, some of which are marked with black arrows in Fig.~\ref{fig:topography}(a). For both graphene on MoTe$_2$ and SnS$_2$ we observe portions of the graphene which are well-adhered to the substrate (flat regions in Figs.~\ref{fig:topography}(c) and (d)), in which a moir\'e pattern is clearly observed. However, the majority of the graphene region is not well-adhered to the substrate, and in these regions we are unable to observe a moir\'e pattern. This may indicate that the MoTe$_2$ and SnS$_2$ crystals are not stable enough in air to support good adhesion with graphene~\cite{Geim2013}. We used the same batch of CVD graphene and transfer method as for other successful devices in this study, helping to rule out bad device fabrication as the source of this poor adhesion. Finally, we find that SnS$_2$ completely evaporates in vacuum above 300 ${^\circ}$C, further suggesting its relative instability.

\begin{figure}[t]
\includegraphics[width=8.5cm]{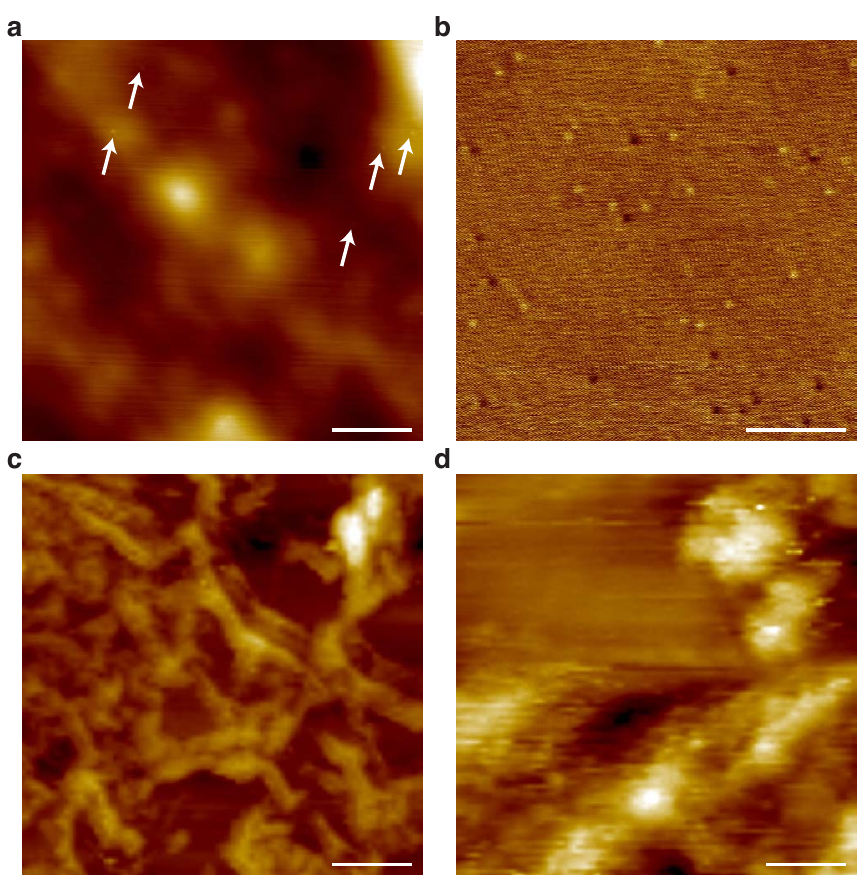} 
\caption{Topography of graphene on (a) WS$_2$, (b) 100 nm thick WSe$_2$, (c) MoTe$_2$, and (d) SnS$_2$. Black arrows in (a) indicate some of the defects. In WSe$_2$, defects are only visible in very thick crystals. The graphene topography in (c) and (d) exhibits flat islands surrounded by very rough regions, suggesting that the adhesion of the graphene to the substrate is not good everywhere. Typical imaging parameters are sample voltages between V$_s$ = 0.05 V and 0.3 V and tunnel currents between I$_t$ = 100 pA and 150 pA. The scale bars are (a) 20 nm, (b), 10 nm, (c), 50 nm, and (d) 10 nm.}
\newpage
\label{fig:topography}
\end{figure}

\section{Defect Topography}

For thin TMD crystals ($<$15 nm), we only observe visible defects buried in the TMD substrates in MoS$_2$ and WS$_2$. However, we also observe defects in graphene on WSe$_2$ (Fig.~\ref{fig:topography}(b)) for a flake about 100 nm thick. Due to band bending of the thick WSe$_2$ substrate~\cite{Larentis2014}, the WSe$_2$ conducts even at small positive gate voltages, whereas the thin WSe$_2$ crystals always remain insulating for the range of gate voltages probed. One possible explanation is that the gate is able to charge defect states in the thick crystals which are not accessible in the thin crystals. This suggests that despite the lack of visible defects in the thin WSe$_2$ samples presented in the main text, there are still localized defect states which can contribute to scattering in the graphene. 

For graphene on MoS$_2$ and WS$_2$ where defects are visible even in thin crystals, we find we can tune the appearance of the defects with sample and gate voltages. For example, Figs.~\ref{fig:defects}(a) and (b) show topography of the same 15 nm region of graphene on MoS$_2$, taken at different sample and gate voltages. There is one large, strong defect in Fig.~\ref{fig:defects}(a), but this defect is not visible in Figs.~\ref{fig:defects}(b) and is instead replaced by many smaller defects in different locations. Similar behavior is observed in graphene on WS$_2$ as well. A full classification of the exact nature of these defects is outside the scope of this work.

\begin{figure}[t]
\includegraphics[width=8.5cm]{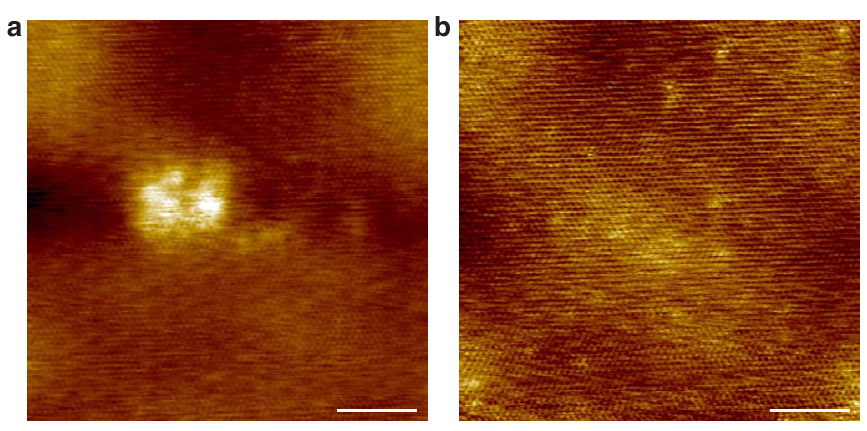} 
\caption{High resolution topography of defects in graphene on MoS$_2$. (a) Topography imaged at V$_s$ = 0.3 V and V$_g$ = 0 V. (b) Topography of the same location at V$_s$ = -0.8 V and V$_g$ = +60 V. There is a single large defect in (a) which disappears in (b) and is replaced by many small defects. I$_t$ = 150 pA and the scale bar is 0.3 nm for both images.}
\newpage
\label{fig:defects}
\end{figure}

\section{Gate Maps of Graphene on TMDs}

\begin{figure}[t]
\includegraphics[width=8.5cm]{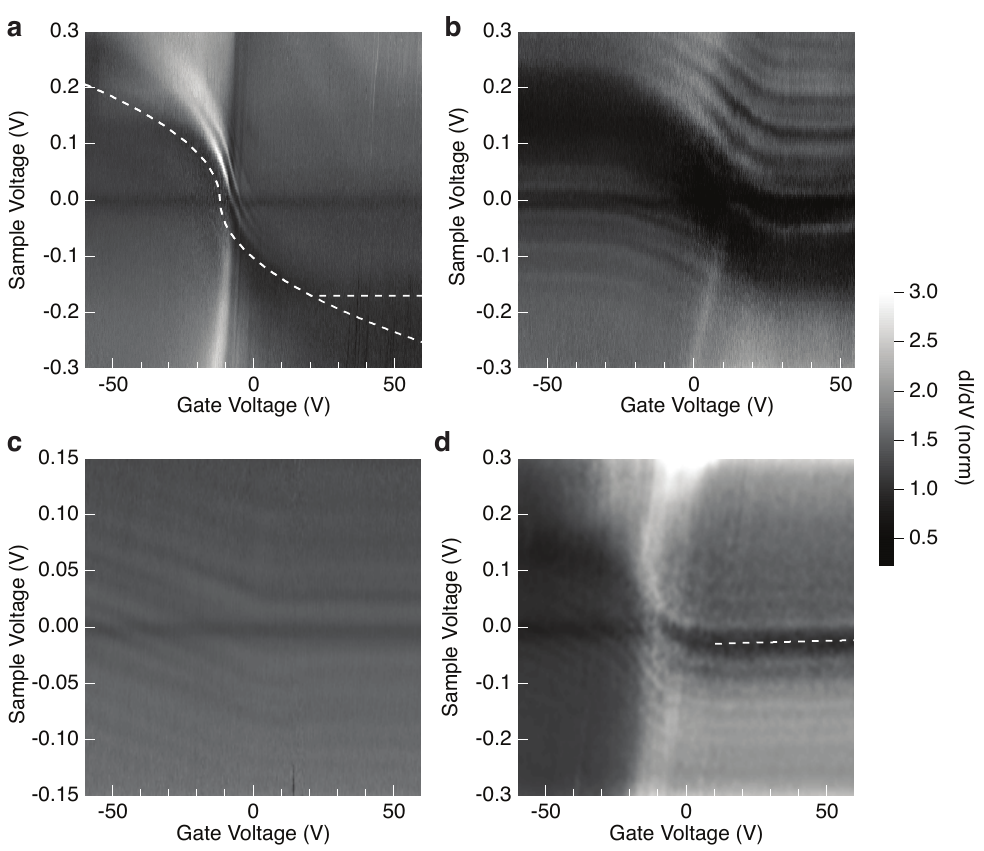} 
\caption{Gate maps of graphene on (a) WS$_2$, (b) MoTe$_2$, (c) SnS$_2$, and (d) MoS$_2$. All exhibit extra resonances surrounding the Dirac point due to intravalley scattering. The white dotted line in (a) represents the position of the Dirac point. The split at V$_g$ $>$ +20 V represents the ambiguity in fitting the Dirac point at large sample voltage. The white dotted line in (d) represents the position of the Dirac point when the MoS$_2$ is conducting. The positive dV$_s$/dV$_g$ indicates the negative compressibility of the sample.}
\newpage
\label{fig:gatesweeps}
\end{figure}

Figures~\ref{fig:gatesweeps}(a) - (c) show gate maps for graphene on WS$_2$, MoTe$_2$ and SnS$_2$ (similar to those shown in Figs. 2(a) and (b) of the main text for MoS$_2$ and WSe$_2$). For graphene on MoTe$_2$ and SnS$_2$, these measurements are taken in well-adhered regions of the sample, as far as possible from the adhesion boundaries. All show the presence of extra resonances surrounding the Dirac point, indicative of intravalley scattering of electrons in graphene. It is difficult to determine if the WS$_2$ crystal becomes conducting at positive gate voltages, as the movement of the Dirac point is expected to slow as the square root of gate voltage, and the energy resolution of the Dirac point becomes worse as it moves further from the Fermi energy. Carriers may populate the WS$_2$ as low as V$_g$ = +20 V, but we are unable to rule out that the WS$_2$ always remains insulating within this range of gate voltage. The white dotted lines on Fig.~\ref{fig:gatesweeps}(a) represent the extremes of these two experimentally indistinguishable positions of the Dirac point (the line which stops moving with gate voltage represents the case where the WS$_2$ becomes conducting). We were unable to obtain global transport in our graphene on WS$_2$ device, so this method could not be used to help resolve the ambiguity. In both graphene on MoTe$_2$ and SnS$_2$, the movement of the Dirac point unambiguously appears to stop above small positive gate voltages.

Figure ~\ref{fig:gatesweeps}(d) shows a second gate map for graphene on MoS$_2$, exhibiting considerably stronger scattering resonances than those of Fig. 2(a) of the main text. The strength of these peaks varies considerably with sample position, and the gate maps chosen for Fig. 2 of the main text exhibited weaker than usual peaks (i.e. the scattering is generally quite strong). The white dashed line at positive gate voltage marks the position of the Dirac point when the MoS$_2$ is conducting. Since the MoS$_2$ screens the gate, there should be no movement of the Dirac point. However, the Dirac point actually shows a slight movement towards more positive sample voltages with increasing gate voltage, which is opposite its normal movement. This positive dV$_s$/dV$_g$ is indicative of the negative compressibility of the graphene on MoS$_2$ heterostructure.

\section{Dependence on Anneal Temperature}

Ref. ~\cite{Kretinin2014} suggests that annealing above 150 $^{\circ}$C has a degradative effect on the mobility of graphene on TMD devices. We have tested this by making graphene on MoS$_2$, WS$_2$, and WSe$_2$ samples with no annealing until the heterostructure was completed, at which point the devices were annealed in vacuum at 150 $^{\circ}$C. We observe qualitatively similar behavior in these devices to those annealed at higher temperatures. Specifically, these devices exhibit a similar density of visible defects, similar intravalley and intervalley scattering states, as well as similar charge fluctuations. As a final test, we further annealed the graphene on WSe$_2$ device to 300 $^{\circ}$C and observed no signatures of degradation. This is in apparent contradiction to the results of Ref. ~\cite{Kretinin2014}. However, as our measurements are local in nature, it is still possible that larger scale rearrangement of trapped dopants is responsible for the degradation of the device mobility.

\newpage
\end{document}